# How is Water released in Hydrogen-Based Metal Oxide Reduction? Unraveling the Kinetic Bottleneck in Sustainable Metal Production

Zhaoxi Chen[#1], Yulu He[#2], Zhuoran Yao[#2, 3], Chunwen Wang[#4], Lei Lei[1], Jun Cai[2], Martin Aaskov Karlsen[5], Claudio Pistidda[6], Wu Zhou[*4], Marc-Georg Willinger[7], Yan Ma[*8], Dierk Raabe[*9], Zhi Liu[1], Zhu-Jun Wang[*2]

[1] Center for Transformative Science (CTS), ShanghaiTech University, No.393 Huaxia Rd, Shanghai 201210, China
[2] School of Physical Science and Technology (SPST) & Shanghai Key Laboratory of High-resolution Electron Microscopy, ShanghaiTech University, No.393 Huaxia Rd, Shanghai 201210, China
[3] Suzhou National Laboratory, No.388 Ruoshui Rd, Suzhou Industrial Park, Jiangsu 215000, China
[4] School of Physical Sciences and CAS Key Laboratory of Vacuum Physics, University of Chinese Academy of Sciences, No.1 Yanqihu East Rd, Beijing 101408, China
[5] Deutsches Elektronen-Synchrotron DESY, Notkestr. 85, 22607 Hamburg, Germany
[6] Institute of Hydrogen Technology, Helmholtz-Zentrum hereon GmbH, D-21502 Geesthacht, Germany
[7] Technical University of Munich, Chair of Electron Microscopy, Lichtenbergstrasse 4, Garching, 85748, Germany
[8] Department of Materials Science & Engineering, Delft University of Technology, 2628 CD Delft, The Netherlands
[9] Max Planck Institute for Sustainable Materials, 40237 Düsseldorf, Germany

## Abstract

Hydrogen-based direct reduction of metal oxides is a ubiquitous solid-gas redox process central to geophysics, sustainable metallurgy, redox energy cycles and catalysis. During this process, hydrogen removes lattice oxygen to form water, yet product water has long been regarded as a passive exhaust, and its nanoscale formation, trapping and removal remain poorly understood. Here, we directly observe redox-product water release from iron oxide during hydrogen-based direct reduction. Because water removal emerges from coupled structural, chemical and crystallographic evolution across multiple length-scales under realistic non-equilibrium reaction-conditions, we establish a correlative multiscale in-situ approach that links pore evolution, molecular water signatures, phase transformation and chemical-state evolution during hematite reduction. We uncover a mechanism in which oxygen removal induces closed nanopores spatially delocalized from reaction surfaces, causing transient trapping of water vapor. Water is released only when these pores coalesce into a percolating network connected to the surface, coinciding with and accelerating the onset of the hematite-to-magnetite transformation. These findings show that dynamically evolving pore topology governs mass transport and redox kinetics in solid-gas reactions, closing a critical mechanistic gap in product-water removal and providing nanoscale guidance for hydrogen-based metal extraction, reactor design, and sustainable redox energy technologies under practical conditions.

[*] Author to whom correspondence should be addressed: wuzhou@ucas.ac.cn; yan.ma@tudelft.nl; d.raabe@mpie.de; wangzhj3@shanghaitech.edu.cn.

## Introduction

The reduction of metal oxides to their metallic forms is one of the most foundational chemical processes in human civilization. From the early metallurgical breakthroughs of the bronze and iron ages to the industrial revolutions that shaped the modern world, this chemical process has enabled the production of structural materials, functional devices, and, more recently, advanced energy technologies[1-5]. At its core, metal oxide reduction involves converting thermodynamically stable oxides into metals or suboxides, typically using chemical reductants such as carbon, carbon monoxide, or hydrogen.

In the industrial context, solid-state reduction underpins the extraction and refining of key metals, including iron, copper, and nickel *etc*[6-9]. These processes have historically relied on the use of fossil-carbon sources as reductants, making the metallurgical sector one of the largest contributors to anthropogenic greenhouse gas emissions, with nearly 10% of global $CO_2$ emissions[1]. In response, hydrogen has gained attention as a sustainable alternative, offering a carbon-free reductant that generates only water as a by-product[9]. This shift promises to decarbonize key sectors of metal production and supports broader efforts toward carbon neutrality[2]. Beyond primary metal extraction, hydrogen-based direct reduction (HyDR) plays an essential role in advanced materials processing and applications, from functional materials for electronics and catalysts, to developing new energy storage and conversion technologies[9]. In materials science and surface engineering, controlled HyDR of metal oxides allows for the tailoring of oxidation states, phase compositions and topology, as well as defect structures, directly influencing properties such as electrical conductivity, magnetism, and catalytic performance[9]. According to thermodynamics, hydrogen can reduce over 30 transition and post-transition metal oxides[10,11], particularly at elevated temperatures, following the simplified reaction $MO_x+xH_2 \rightarrow M+xH_2O$ (where M stands for a metal element), **Figure 1**a.

This reaction involves lattice oxygen in metal oxides ($MO_x$) reacting with dissociated hydrogen atoms (H) to form water ($H_2O$) and create oxygen vacancies in the oxides[12,13]. These vacancies subsequently diffuse through the oxide lattice, aggregating to form nanoscale voids[14], **Figure 1**c. With progressing reduction, this process leads to the development of interconnected pores and finally to three-dimensional pore networks within the oxide bulk, which significantly influence reduction kinetics[15,16]. The closed nanovoids and pores potentially trap water, lowering the local partial pressure of hydrogen (the thermodynamic driving force), thus hindering reactions and even promoting re-oxidation of the reduced metallic phase[14]. Conversely, the newly generated connected pore networks facilitate mass transport, enabling both inbound hydrogen diffusion and outbound water diffusion, thereby accelerating the reaction. Despite substantial advances in understanding the thermodynamics and transport processes underlying HyDR of metal oxides, two fundamental mechanistic questions remain unresolved: Where does water form in the highly nanoporous and partially reduced oxide matrix and how is water evacuated and removed from the reaction system?

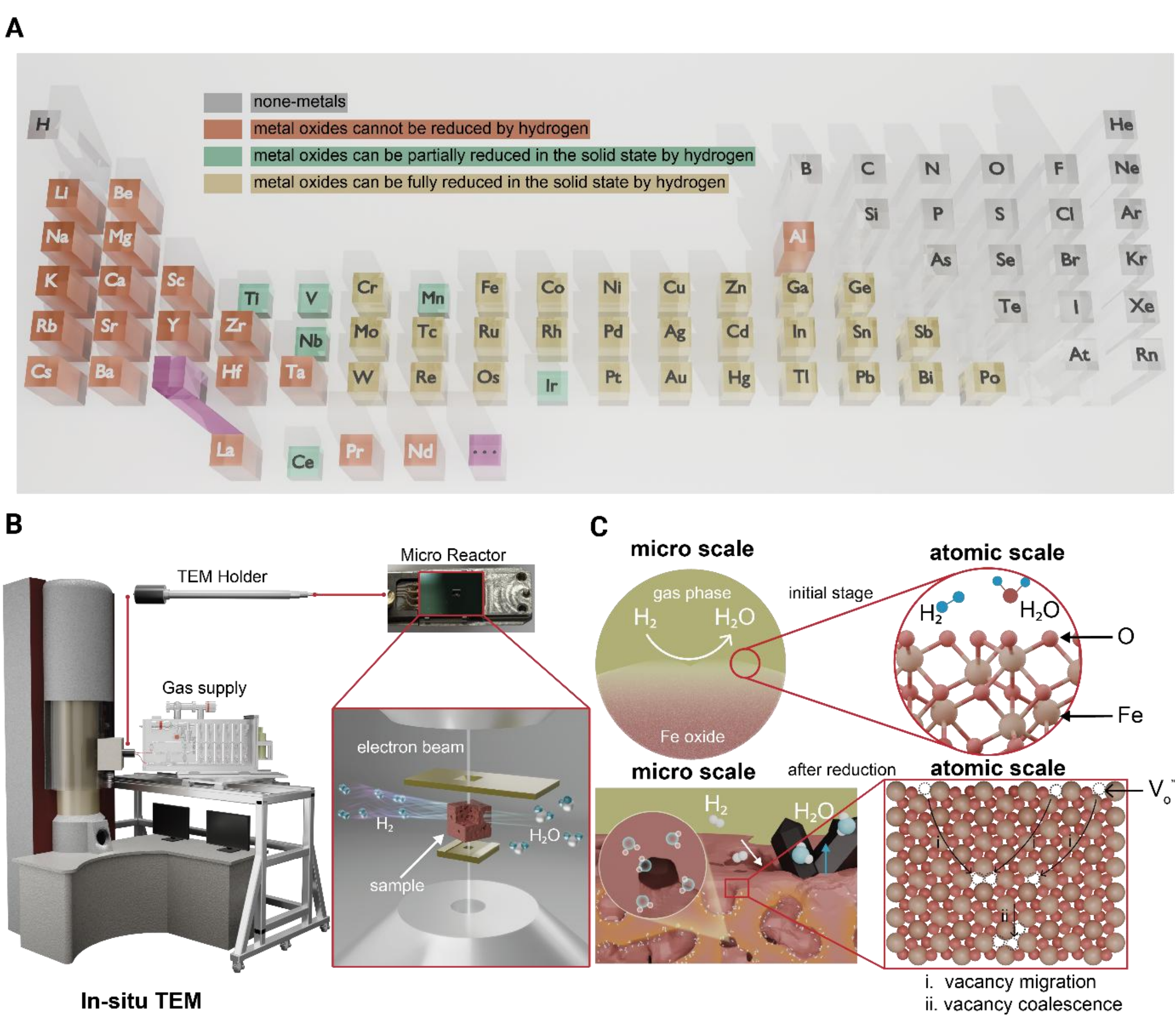

***Figure 1**. **Schematic overview of hydrogen-based direct reduction (HyDR) of metal oxides.** (a) Periodic table overview of the thermodynamic reducibility of metal oxides (in the solid state) in hydrogen. Colors indicate whether the selected oxide of each element is predicted to be reducible to the elemental metal, reducible only to a lower/suboxide, or not reducible under the specific conditions. The classification was evaluated for $P_{tot}$ = 1 bar. The relevant thermodynamic driving force depends on temperature and the imposed $P_{H2O}/P_{H2}$ ratio, with feasibility assessed using Ellingham-type calculations; therefore, the map should not be interpreted as an absolute property of the elements, and kinetic limitations are not considered. Specific applications in metallurgy, energy, catalysis, and sensing are detailed in Table S1. (b) Experimental setup for in-situ observation of HyDR using gas-cell transmission electron microscopy (TEM). (c) Multiscale illustration of the reduction mechanism. The panels depict the transition from the initial surface interaction to the post-reduction state, highlighting oxygen vacancy ( $V_O^{\cdot\cdot}$) formation, migration, and coalescence at the atomic scale.*

To answer these questions, we present here the first direct *in-situ* transmission electron microscopy (TEM) observations of porosity evolution and water dynamics, including water release from reduced material exemplarily during HyDR of α-$Fe_2O_3$ nanoparticles (**Figure 1**b). Focus was placed on previously unexplored dynamic evolution of nanovoids and porosity below the reaction surfaces and the associated water trapping and release within and from these nanopores into the gas phase. Due to the highly dynamic nature of these processes, advanced *in-situ* TEM probing was employed, enabling us to prove actual encapsulated water release for the first time directly during the ongoing redox process.

Understanding the removal of water, the redox product of the reduction reaction, is crucial for sustaining the required reduction thermodynamic and kinetic conditions, as trapped water can lower the local driving force and block active sites for the reduction reaction[14,17]. These *in-situ* observations thus close a major knowledge gap by elucidating the dynamic interplay between porosity formation, water trapping and release, and phase transformation, which govern the overall reaction kinetics and are thus essential to understand for designing reactors and feedstocks to turn the 2-billion-ton annual steel sector more sustainable.

## Results and Discussions

### Delocalized Pore Formation Below the Surfaces: Closed Pores

Using *in-situ* TEM techniques, we track the highly dynamic formation and evolution of nanopores and the kinetic behavior of water molecules in real time in a model system of α-$Fe_2O_3$ (hematite). α-$Fe_2O_3$ nanoparticles with an average crystallite size of 148±15 nm (estimated from XRD) are subjected to HyDR at 1 bar in a 10$H_2$-90Ar (in vol%) gas mixture. **Figure 2a-c** presents the three-dimensional distribution of nanoscale voids and pores generated during the HyDR process in partially reduced α-$Fe_2O_3$ samples, as imaged by transmission electron tomography. A deep-learning approach, WBP-UNET, was employed to simultaneously address denoising and missing-wedge artifacts during reconstruction[18]. The α-$Fe_2O_3$ nanoparticles were heated to 573K and held under the reducing atmosphere for 1800s. High-resolution imaging (**Figure 2d**) confirms that the nanoparticles retain the α-$Fe_2O_3$ phase, indicating that the transformation to $Fe_3O_4$ (magnetite) does not occur at this stage. This phase assignment is further supported by SAED analysis presented in a later section. However, nanovoids ($\leqslant$ 5 nm) and internal nanopores (5-50 nm) begin to form within the hematite nanoparticles at this early stage, as evidenced by the projection and sliced voxel images from electron tomography reconstructions, **Figure 2**c. These nanovoids are fully encapsulated by the nanoparticle matrix and are randomly distributed.

To quantitatively characterize the temperature-dependent topological transitions of the internal pore networks, quasi-*in-situ* 3D electron tomography reconstructions were systematically carried out (**Figure 2**e–2g). The rendered three-dimensional volumes alongside their corresponding cross-sectional orthoslices (**Figure 2**e–2g, insets) visually confirm the progressive physical evolution of the sample from its pristine, dense configuration to a highly porous microstructure under elevated reducing temperatures. Crucially, the upper-right inset associated with each tomographic slice explicitly defines its exact spatial positioning within the individual nanoparticle matrix. This spatial correlation provides compelling evidence that pore nucleation preferentially propagates delocalized deep within the bulk interior rather than being confined to the surface atomic layers.

To translate these discrete microstructural profiles into quantitative statistical descriptors, pore diameter distribution histograms are extracted from the tomographic reconstruction datasets (**Figure 2**h-j). At the initial pristine stage (298 K), only a very limited number of pores is detected, indicating the near absence of internal nanoscale voids in the untreated α-$Fe_2O_3$ nanoparticles. Upon elevating the reduction temperature to 573 K and 673 K, the cumulative pore volume increases markedly, as reflected by the substantially higher counts in the diameter distribution.

Increasing the reduction temperature to 673K promotes the formation of non-faceted, near-spherical nanopores within α-$Fe_2O_3$ nanoparticles (**Figure 2**k). The porosity evolution was evaluated by quantitative segmentations of multiple tomography reconstructions. The volume fraction of these closed voids/pores shows only a modest change, from 1.7±0.2% at 573K to 1.0±0.1% at 673K, while their mean diameter increases from 3.1±0.2 nm to 4.3±0.6 nm. Interestingly, the nanovoids become more regularly distributed throughout the nanoparticles, but they are not localized near the surface atomic layers where oxygen removal by hydrogen occurs directly through surface redox reactions. This observation suggests delocalized nanopore formation below the surface, from accumulating point defects left behind by removed oxygen.

The dynamic evolution in crystallography, morphology, and local chemical states of the nanoparticles was further probed in real time using *in-situ* scanning transmission electron microscopy (STEM). Time-lapse imaging reveals highly dynamic nanopore growth via nanovoid coalescence, forming larger spherical pores (≥5 nm), **Figure 2**l-m, prior to the bulk phase transformation from hematite into magnetite. This finding is consistent with *ex-situ* tomographic reconstruction results (**Figure 2a-c**). *In-situ* STEM at elevated temperatures (**Figure 2**k) demonstrates temperature-dependent pore growth, with large spherical nanopores (10-30 nm) dominating at 673K. The consistency between *in-situ* and *quasi-in-situ* analyses confirms that the delocalized nanopore growth is an inherent feature of HyDR[19], originating from the surface reactions and generating closed nanopores within the precursor matrix (here $\alpha$-$Fe_2O_3$).

The formation and evolution of closed nanovoids (<5 nm) and nanopores (5-50 nm) inside the untransformed hematite differ from the pore formation at the $\alpha$-$Fe_2O_3/Fe_3O_4$ interface, and the latter remains confined to $Fe_3O_4$ [15]. Anion oxygen vacancies ( $V_O^{\cdot\cdot}$ ) form at the hematite surfaces due to oxygen loss. These vacancies are then occupied by sub-surface oxygen, as supported by MD simulations[20]. The resulting chemical potential gradient drives oxygen diffusion from the interior towards the surface, simultaneously promoting vacancy migration toward the nanoparticle core[21]. This process results in vacancy clustering, experimentally observed as nanovoids (<5 nm). Subsequently, coalescence of nanovoids into nanopores (5-50 nm) minimizes surface energy (**Figure 2**m-n). These findings highlight delocalized volumetric interactions within the $\alpha$-$Fe_2O_3$ matrix, and the role of these closed pores in HyDR is elucidated next.

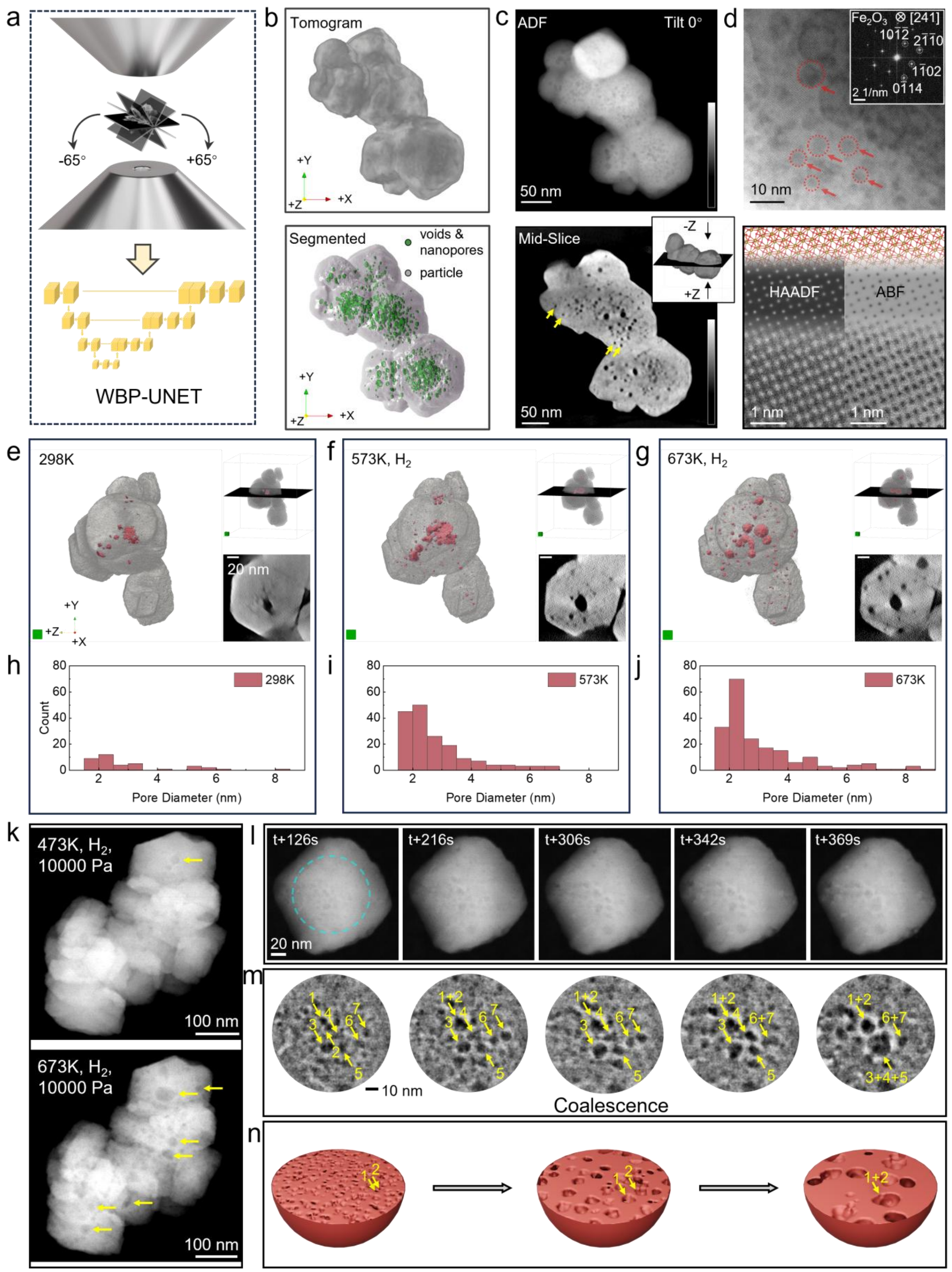

***Figure 2. Coalescence of nanovoids into closed pores inside hematite nanoparticles during HyDR.*** *(a) Schematics of the electron tomographic reconstruction. (b) Electron tomographic images showing nanovoid growth inside the hematite precursor matrix. The top image shows composite view of the tomogram. The bottom image shows segmented particle/pore structure. The samples were obtained after treatment at 573K for 1800 s under a gas mixture of 10 vol.% $H_2$ and 90 vol.% Ar at 1 bar. The corresponding STEM image at zero-tilt angle is shown in (c)-top. The intermediate slice in (c)-bottom at the indicated depth showing internal nanovoid*

*formation with dark contrast to the solid part. (d) High-resolution STEM images showing the nanovoid formation within an α-$Fe_2O_3$ nanoparticle. The upper panel shows the nanoparticle viewed along the α-$Fe_2O_3$-[241] zone axis. The lower panels present the corresponding atomic model, simulated HAADF/ABF images, and experimental HAADF/ABF images along the [241] zone axis, shown as magnified views of the region in the upper panel. Arrows in (c, d) denote representative nanoscale voids ($\lesssim$ ~5 nm). (e-g) Quasi in-situ three-dimensional electron tomography reconstructions under 298K, 573K and 673K. The top panels showcase the reconstructed 3D volumes and their corresponding cross-sectional slices. The upper-right inset indicates the spatial position of each slice within the particle. The green cube denotes a 20 nm × 20 nm × 20 nm volume. (h-j) Quantitative statistical analysis of pore diameter distribution derived from the tomographic reconstruction datasets. (k) In-situ STEM images showing the nanopore formation at the α-$Fe_2O_3$ NPs during HyDR. Images were acquired after 1800 s of treatment at each temperature. The sample was not illuminated between 473 and 673K to minimize the electron beam effect. (l-m) In-situ STEM images showing the coalescence of nanovoids (marked by yellow arrows) at 573 K. (n) Schematics of nanovoid growth through coalescence.*

## Direct Observation of Water Trapping in Closed Pores

Theoretical studies suggest that hydroxyl groups ($(\mathrm{OH})_\mathrm{O}^{\cdot}$) and hydridic hydrogen at oxygen vacancies ($\mathrm{H}_\mathrm{O}^{\cdot}$) mediate bulk reaction during HyDR[22-26]. Adsorbed hydrogen atoms at surface active sites can diffuse into the oxide lattice, producing water ($2(\mathrm{OH})_\mathrm{O}^{\cdot} \rightarrow \mathrm{O}_\mathrm{O}^{\mathrm{x}} + \mathrm{V}_\mathrm{O}^{\cdot\cdot} + \mathrm{H_2O}$) at nanovoid or nanopore surfaces. This water may subsequently become trapped within closed pores, though direct experimental evidence has remained elusive.

To address this knowledge gap, we employed vibrational aloof-beam electron energy loss spectroscopy (EELS)[27] coupled with secondary electron (SE) imaging which resolves the surface morphology and surface pores, if existed, of α-$Fe_2O_3$ nanoparticle. The tomographic reconstruction was first used to map the spatial distribution of internal and surface pores, **Figure 3**a-b, while vibrational EELS was used to probe molecular signatures within the untransformed hematite. **Figure 3**c shows an α-$Fe_2O_3$ nanoparticle with internal nanovoids, also revealed in the SE image. The non-uniform features appearing in the STEM images are indicative of nanovoids within the nanoparticle, but they were not observed on the nanoparticle surface (in SE mode). It is worth noting that the specimen was annealed at 393K for 8 hours in high vacuum ($\lesssim 10^{-3}$ Pa), to remove adsorbed species on the nanoparticle surfaces and from open pores prior to spectroscopy, ensuring that the detected signals originated solely from internal nanovoids or nanopores, not from contaminated open surfaces.

Vibrational EELS in aloof-beam mode identified the molecular motifs within the untransformed hematite matrix, **Figure 3**f-h. Hydroxides exhibit a sharp peak around 450 meV, whereas hydrates and water display broader peaks centered at about 430 meV[28]. Vibrational spectra acquired near closed pores (**Figure 3**c) revealed clear OH stretching modes matching molecular $H_2O$ (**Figure 3**f), confirming water trapping within the closed pores. Notably, these vibrational EELS signals persisted up to ~100 nm from the specimens, **Figure 3**e, h, decaying exponentially with a factor of ~ $1.5 \times 10^{-3}$ Å$^{-1}$, in agreement with prior reports[29]. These findings suggest that the delocalized nanovoids in partially reduced and untransformed α-$Fe_2O_3$ nanoparticles function as *in-situ* water reservoirs. Such trapped water elevates the local partial pressure of water ($P_{H_2O}/(P_{H_2O} + P_{H_2})$), potentially impeding the reaction kinetics and compromising process efficiency[14].

Next, we examine water release associated with pore coalescence in real space. **Figure 3**d shows a representative surface-connected pore approximately 10 nm in diameter, visible in both the STEM and SE images. Vibrational EELS detects no molecular $H_2O$ signal near these interconnected surface pores, indicating that water is released once internal nanovoids coalesce and connect to the nanoparticle surface. This observation contrasts with the clear OH-stretching signals observed at closed pores, as shown in **Figure 3**c, f. We therefore further investigate how water is removed from the reaction system through dynamic pore-connection pathways.

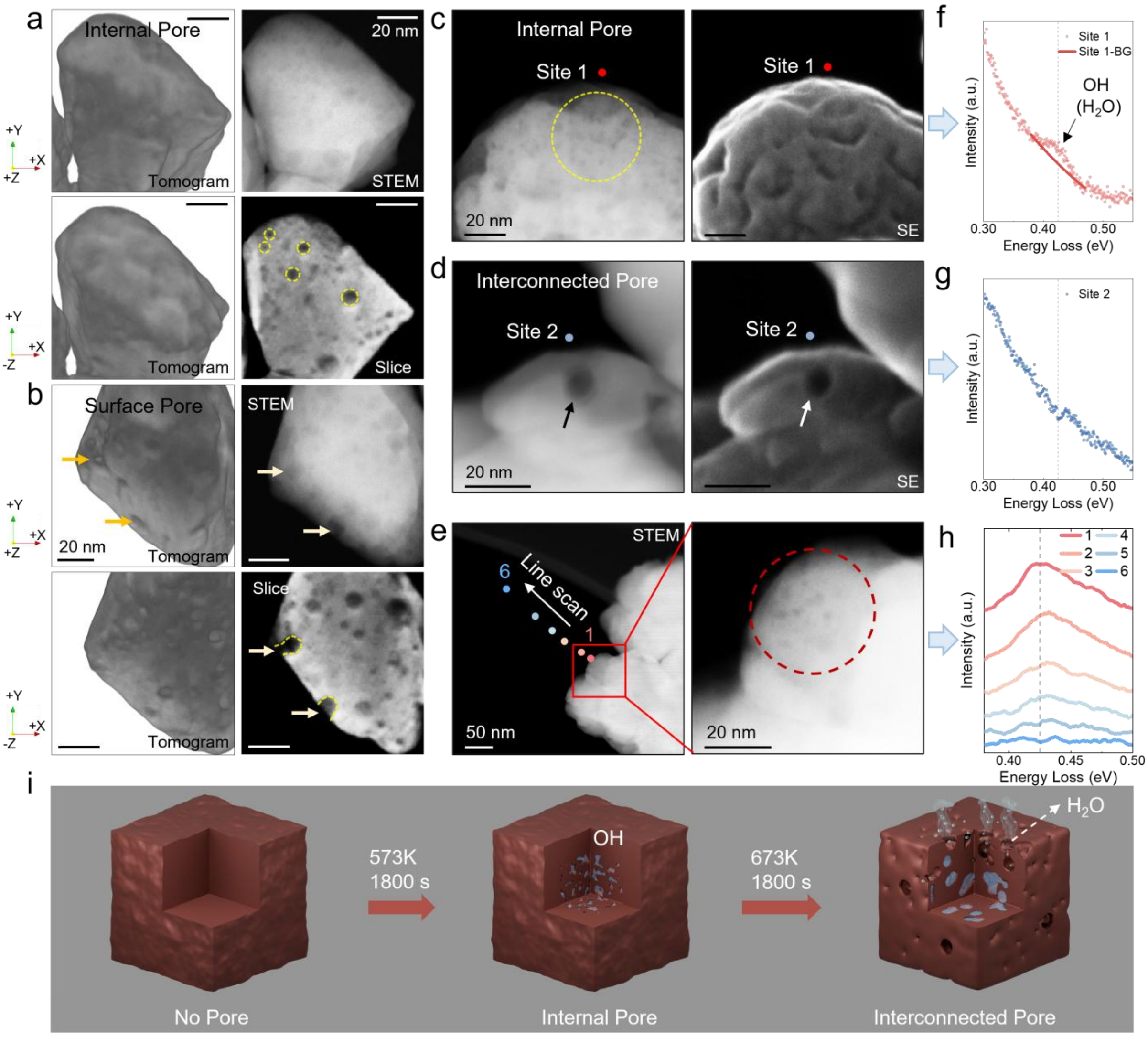

***Figure 3. Experimental proof for water storage in closed pores.*** *(a-b) STEM tomography of hematite nanoparticles. Left panels show reconstructed 3D volumes viewed along the electron-beam axis (+Z and horizontally flipped -Z axes) to highlight surface morphology. Right panels show corresponding STEM images and representative tomographic slices, distinguishing internal pores (enclosed circles) from surface-connected pores (arrows). (c-e) STEM and secondary electron (SE) images showing the nanopores in hematite nanoparticles. In (c), small internal pores (circles) are visible in STEM but absent in the SE image, confirming that they are enclosed within the particle. In (d), a large pore (arrow) is visible in the SE image, indicating its connection to the particle surface through pore coalescence. (f-h) Aloof-mode vibrational EELS spectra acquired at positions marked in (c-e). A distinct peak at around 430 meV is assigned to the -OH stretching mode. The specimen was degassed prior to analysis to remove adsorbed species. Comparison of (f) and (g) shows that the -OH signal is present at closed internal pores but disappears at surface-connected pores. The hump around 0.43 eV in (f) corresponds to the signal from hydrates/water stored in the internal closed pore, while the spectrum in (g) suggests hydrates/water released through interconnected pores. (i) Schematic illustrating water trapped within internal porosity and their removal via interconnected surface channels during HyDR of hematite.*

## Dynamics of Pores and Water Release

To elucidate the dynamics of the porosity evolution and its associated topological and percolative features, we systematically investigated nanopore growth and connectivity an *in-* *situ*. These investigations were conducted at elevated reduction temperatures equal or exceeding 673 K, as temperature dominantly governs nanopore evolution in hematite nanoparticles during HyDR[5,30,31]. Above this temperature, a critical transition occurs, where nanopores undergo continuous growth, ultimately traversing the particle structure to connect with either the oxide particles' surfaces or with adjacent pores, **Figure 4**. The growth trajectory of a single nanopore was quantitatively analyzed through electron-scattering line profile analyses (**Figure 4**a, b, and f) and lattice-indexed measurements (**Figure 4**d-e, and g), revealing anisotropic expansion preferentially elongated along the $\{30\bar{3}0\}$ normal-plane directions of hematite (**Figure 4**c).

This rapid anisotropic growth facilitated connection to the particle surface, substantially altering reduction kinetics. Upon pore breaching to the surface, the internal pore surface area became directly accessible to the reducing gas atmospheric conditions, while otherwise the hydrogen must diffuse in atomic form through the solid after the dissociation of $H_2$ on the particle surface. Consequently, the elevated hydrogen diffusion coefficient at the gas-solid interface and enhanced water vapor removal via connected porosity induce a transition in the rate-limiting step[2,32]. This configuration promotes heterogeneous reduction reactions throughout the interconnected pore network, after which the reaction rate becomes constrained by solid-state phase boundary kinetics, *i.e.*, the interfacial reaction governing the phase transformation from hematite to magnetite. This mechanistic shift also rationalizes the observed pore morphology transition: internal spherical pores arise from vacancy aggregation and void coalescence, whereas interconnected non-regular pores are further governed by solid-gas interactions at the pore surface.

High-temperature *in-situ* observations (at 673 K) captured the dynamic migration and coalescence of multiple nanopores, **Figure 4**h-j. The formation of interconnected pore networks was quantitatively demonstrated through time-stacked area analysis (**Figure 4**i-j) and confirmed by *ex-situ* tomography (**Figure 4**k-l). This interconnection process occurs within a timeframe of $10^2$ s with a pore growth rate on the order of $2 \times 10^{-2}$ nm/s. These 3D reconstructions revealed the development of continuous pathways linking internal nanovoids with the particle surface. Surface-initiated pores progressively extended into the nanoparticle interior, establishing the interconnected porosity network, which leverages faster mass transport during HyDR, specifically outbound water transport. Such observations of the very fast nanopore evolution pathway provide critical insights into the dynamic behavior of nanopore formation and evolution during HyDR.

These findings demonstrate that water can accumulate within the closed pores during HyDR and how it is subsequently released through a dynamic interconnection process involving steps: (i) internal void coalescence, (ii) pore interconnection, (iii) surface breakthrough, and (iv) water desorption. The associated timescales of these individual processes under our operational conditions are of the order of $10^3$ s for step (i), $10^2$ s for steps (ii) and (iii). This highly dynamic coupling between nanopore structure evolution and water trapping/release highlights the critical role of nanoscale mass transport pathways in determining the overall reduction efficiency of hematite nanoparticles during HyDR.

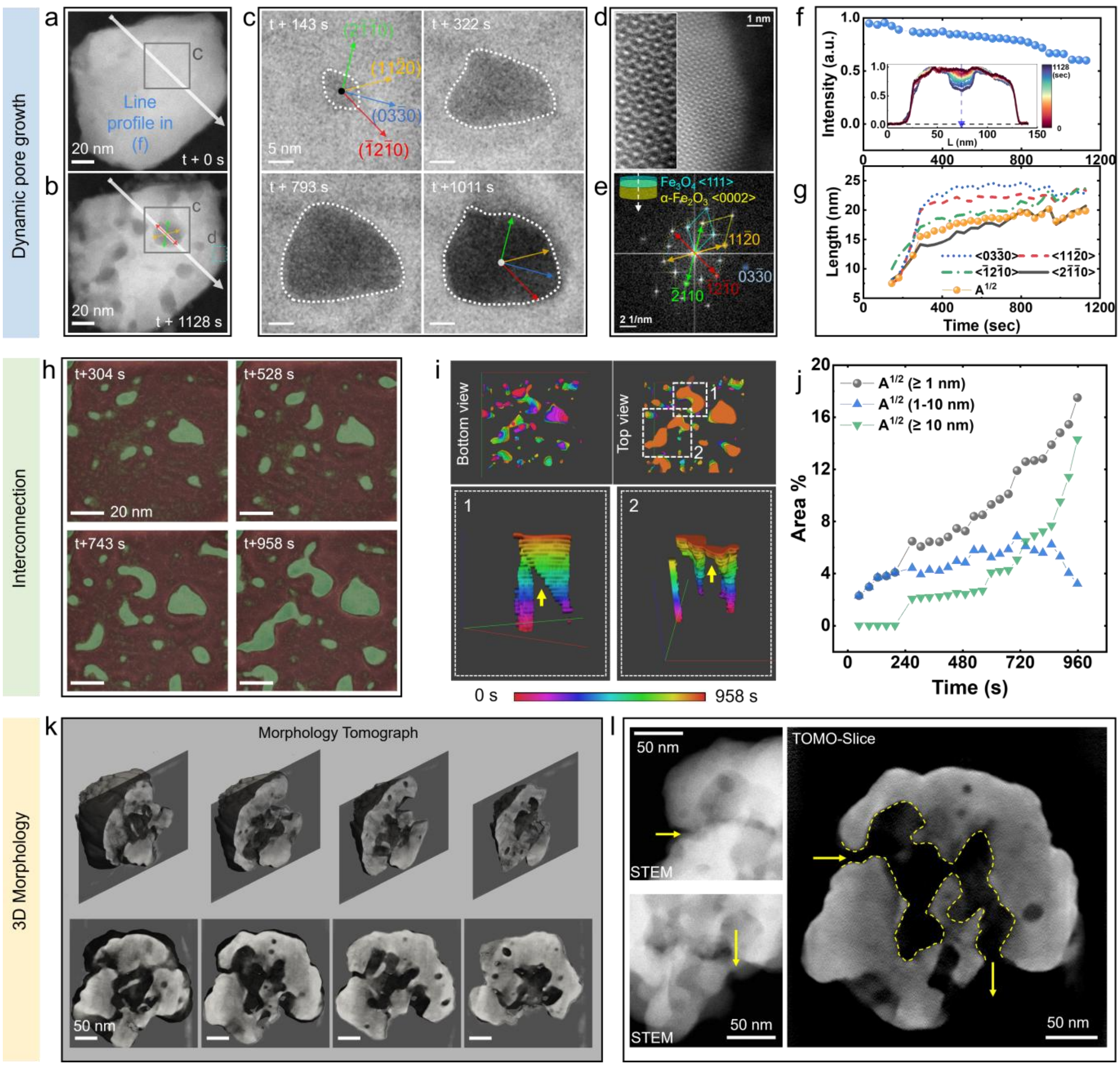

*__Figure 4. Dynamic pore growth and free surface augmentation during HyDR of__ __$\alpha$-$Fe_2O_3$__. The observation is made on nanopores traversed to the hematite surface view along the crystallographic <0002> axis during HyDR at 673K. (a-b) The morphological change of the hematite nanoparticle at t=0 s and t=1128 s. Note the pore edge extending to the hematite surface. (c) Time-lapse STEM images showing the growth of a nanopore. (d) High-resolution image showing a layered structure of hematite and magnetite after the reduction. (e) The*

*Fourier transform of (d) showing two sets of diffraction results for hematite and magnetite. (f) The normalized electron scattering intensities across the pore marked in (a-b). The vertical axis is normalized by the difference between the maximum intensity on the particle site and the background. (g) The extracted length profile in normal directions with specific crystallographic planes. The crystallographic orientations are indexed in (e) and marked in (b) and (c). (h) Time-lapse images showing the movement of multiple nanopores during hydrogen reduction at 673K. Nanopores (highlighted in green) were identified using a random-forest segmentation algorithm. (i) Composite 3D perspective visualization of nanopore evolution. Arrows indicate points where coalescence leads to interconnection between nanopores. Bottom/Top view: Both perspectives are aligned with the time axis: the bottom view progresses chronologically front-to-back, while the top view progresses back-to-front. (j) The evolution of the characteristic pore size and projected area over time quantified from (h). (k) 3D tomographic images showcase a hematite particle that has been subjected to reduction at 673K for a duration of 3600 s. (l) Comparison of STEM and tomographic images depicting porous channels connecting to the particle surface. The highlighted region represents a connected porous zone formed inside the hematite particle. The arrow marks the pore connected to the outer surface.*

## Spontaneous Magnetite Growth with Surface Pore Connection and Water Release

During HyDR of $\alpha$-$Fe_2O_3$, magnetite nucleation occurs preferentially at surface defect sites[33]. High-resolution imaging confirms magnetite nucleation near porous hematite particle edges (**Figure 4**d-e). This process occurs within a timeframe of 10-$10^2$ s. Spontaneous magnetite nucleation initiates when nanopores breach the nanoparticle surface, inducing concurrent morphological transformations in both hematite and magnetite (**Figure 5**). Such a stage sustains within a timeframe of $10^3$ s until the parent hematite phase is entirely consumed. Interconnected porosity enables volumetric interactions that propagate through sub-surface layers at the $\alpha$-$Fe_2O_3$/$Fe_3O_4$ interface. This facilitates the phase transformation pathway, as also underpinned by *in-situ* synchrotron X-ray diffraction experiments and complementary morphological analyses, in line with earlier experiments[3,30,31]. The observed morphological evolution directly correlates magnetite formation with nanopore growth, connection, and – most importantly - with the associated water release.

These findings establish the spatiotemporal coupling between nanopore evolution and magnetite formation dynamics. At reduction temperatures below 673K, isolated spherical nanovoids form primarily within hematite particles. At temperatures $\geqslant$673K, however, inner porosity breaches the particle surface, providing preferential nucleation sites for magnetite. Crucially, this interconnected pore network provides pathways for expelling water vapor generated during reduction. The efficient removal of water promotes continued hematite reduction at the breach sites, directly facilitating magnetite nucleation and growth. As reduction progresses, these surface-connected pores advance deeper into the hematite precursor, further integrating the internal nanovoid structure. Pre-existing internal voids/pores significantly shorten diffusion pathways required for percolation to the surface. Consequently, the pore fraction connected to the surface increases substantially during the hematite-to-magnetite transition, enabling stepwise improved percolative conditions during surface restructuring.

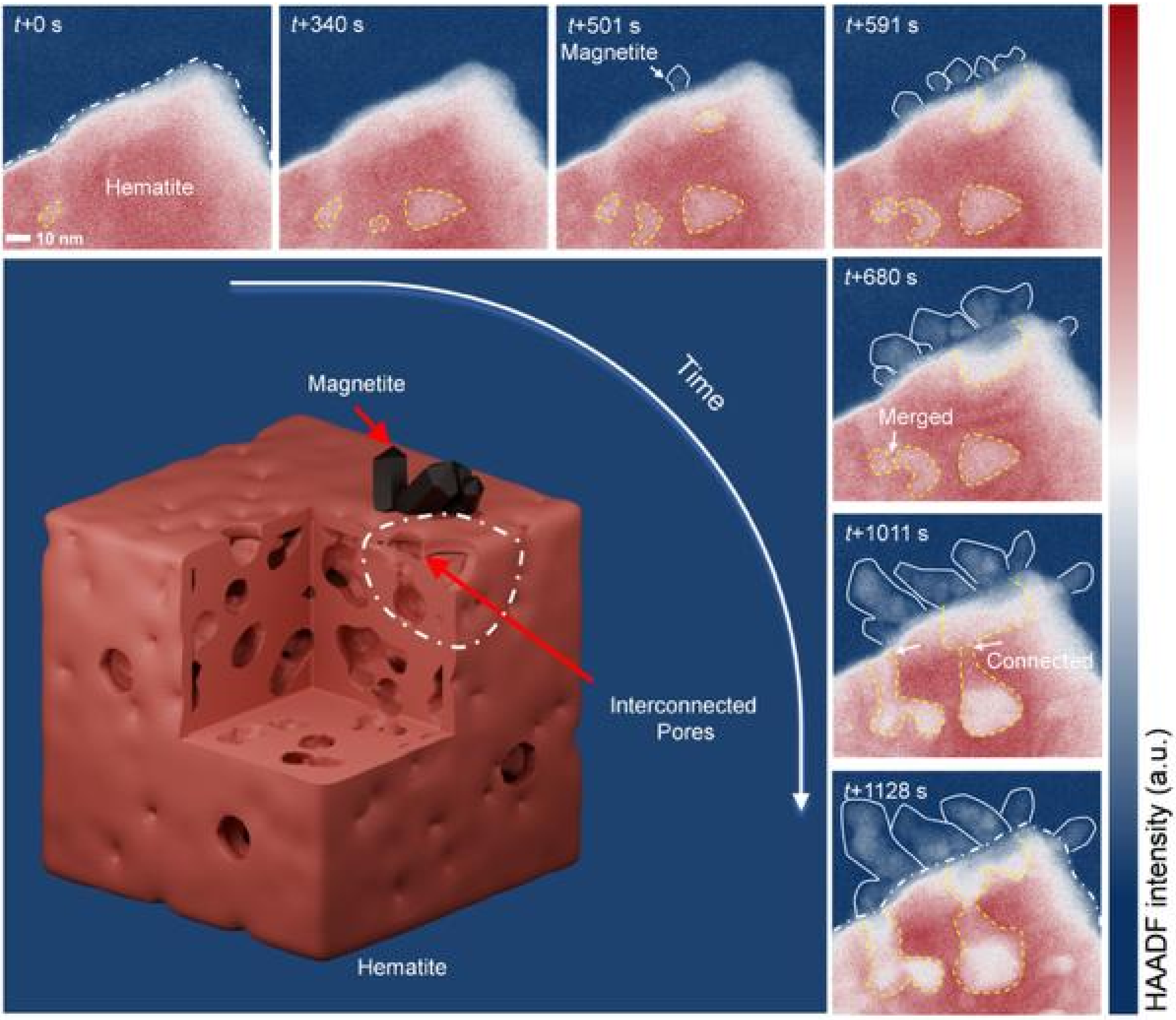

***Figure 5. Spontaneous open-pore and magnetite growth during the first reduction step (α-$Fe_2O_3$ → $Fe_3O_4$).*** *The observation was conducted under a gas mixture (10 vol.% H2 and 90 vol.% Ar) at 673K, 1 bar. The images were acquired under HAADF-STEM mode. A dash-dot line marks the initial perimetry of the hematite NP precursor. The yellow dashed lines outline the free volumes (porosity) at the hematite site. The solid lines associated with the magnetite nucleation are used for guiding the eye, suggestive of magnetite nucleation on the surface of hematite and protruding out.*

## Conclusions and outlook

We present the first direct, atomic-scale observation of redox product water release from a metal oxide surface during hydrogen-based direct reduction, revealing how dynamic mass transport governs oxide reduction kinetics. Using *in-situ* transmission electron microscopy combined with three-dimensional electron tomography and vibrational electron energy-loss spectroscopy, we directly capture the formation and evolution of nanopores in hematite, the transient trapping and subsequent release of water vapor, and the concurrent phase transformation to magnetite. During the early stages of reduction, oxygen removal from the surface lattice generates vacancies that migrate and cluster into nanovoids and closed nanopores within the hematite matrix, spatially decoupled from active reaction surfaces. Notably, these delocalized nanopores form before the appearance of magnetite, indicating an intrinsic early-stage pore formation mechanism.

As reduction proceeds above 673 K, isolated nanopores coalesce into a percolating network that markedly enhances gas transport, particularly the removal of product water, thereby accelerating the hematite-to-magnetite transformation. These findings demonstrate that dynamically evolving nanopores, initially acting as water-trapping sites before transitioning into connected transport pathways, are not passive by-products but active enablers of efficient hydrogen reduction. By directly resolving nanoscale water transport dynamics, this work identifies a previously unrecognized kinetic bottleneck in solid-gas redox reactions and informs the rational design of advanced metallurgical processes, catalytic redox systems, and hydrogen-based energy technologies.

**Author Contributions**

Z-J.W. and Z.C. conceived this project and supervised the research. Experiments and analysis of *in-situ* TEM and *in-situ* SAED were conducted by Z.C. and Y.H. Experiments and analysis of Tomography and quasi *in-situ* tomography were conducted by Z.C., Y.H. and L.L.. Experiments and analysis of Aloof vibrational EELS were conducted by W.Z. and C.W.. Experiments and analysis of *in-situ* XRD were conducted by D.R., Y. M., M. A. K. and C. P.. Experiments and analysis of *in-situ* NAP-XPS were conducted by J.C. and Z.Y.. Z. C., Y. H., Z. Y., and C. W. contributed equally to this work. Important contributions to the interpretation of the results, conception and writing of the paper were made by Z.C. and Z.-J.W. All authors participated in the scientific discussion.